# TCDRM: A Tenant Budget-Aware Data Replication Framework for Multi-Cloud Computing

Santatra Hagamalala Bernardin [1], Riad Mokadem [2], Franck Morvan [2], Hasinarivo Ramanana[1],

Hasimandimby Rakotoarivelo [1]

[1] Université of Antananarivo, Mention Informatique et Technologies, Madagascar

[2] Institut de Recherche en Informatique de Toulouse (IRIT), University of Toulouse, Toulouse, France

**Abstract**. Multi-cloud computing systems face significant challenges in ensuring acceptable performance while adhering to tenant budget requirements. This paper proposes a tenant budget-aware (tenant-centric) data replication framework for Multi-Cloud Computing (TCDRM). The proposed strategy dynamically creates data replicas based on predefined thresholds for response time, economic budget of the tenant and data popularity. TCDRM employs a heuristic replica placement algorithm that leverages the diverse pricing structures of multiple cloud providers. The TCDRM strategy aims to maintain the required performance without exceeding the tenant's budget by taking advantage of the capabilities offered by multi-cloud environments. The middleware considered acts as an intermediary between tenants and multiple cloud providers, facilitating intelligent replica placement decisions. To achieve this, the proposed TCDRM strategy defines strict thresholds for tenant budget and response time. A performance evaluation is conducted to validate the effectiveness of the strategy. The results show that our approach effectively meets tenant performance objectives while respecting their economic constraints. Bandwidth consumption is reduced by up to 78% compared to non-replicated approaches, and average response time for complex queries is decreased by 51%, all while adhering to tenant budget limitations.

## 1. Introduction

Multi-cloud environments have emerged as a preferred paradigm for deploying data-intensive applications due to their ability to combine resources from multiple cloud providers. These environments offer flexibility, fault tolerance, and cost-optimization opportunities by allowing tenants to select providers based on pricing (Armbrust et al., 2010), performance and geographic availability (Stantchev and Schropfer, 2009). However, they also introduce significant challenges, particularly in the context of data replication, where economic and performance factors must be balanced across heterogeneous infrastructures.

One of the most pressing challenges in multi-cloud data management is enabling effective data replication that considers not only technical performance (e.g., latency, availability), but also the budgetary constraints of individual tenants. In traditional replication strategies, the focus is often placed on maximizing provider profit or system-wide efficiency, with limited consideration of tenant-specific cost sensitivity (Mokadem and Hameurlain, 2020; Khelifa et al., 2020). This leads to inefficiencies, especially in multi-tenant scenarios where each user may operate under different budget and SLA expectations.

While several replication strategies have been proposed for cloud environments (Milani and Navimipour, 2016; Shakarami et al., 2021; Mokadem et al.,2022; Dugyani and Govardhan, 2024), most of them either assume provider-oriented or are tailored for single-cloud contexts. A few studies have considered tenant-oriented approaches (Limam et al., 2019), but these are rarely extended to support multi-cloud systems or cost-aware decision-making that reflects the pricing diversity of different providers. This results in a significant gap in the literature: how to design a data replication strategy that adapts to tenant budgets in a multi-cloud environment.

In this paper, we address this gap by proposing a tenant budget-aware (tenant-centric) data replication Framework for Multi-Cloud Computing (TCDRM). The proposed data replication strategy prioritizes the tenant's budget while ensuring acceptable query performance. TCDRM dynamically analyzes execution cost and performance metrics after each query execution and evaluates whether the creation of new replicas is necessary based on predefined thresholds in the SLA and tenant-defined cost ceilings. When these thresholds are violated, TCDRM triggers a decision-making process to analyze query dependencies and identify candidate data for replication. A cost-aware replica placement heuristic is then applied to select the most suitable cloud provider, taking into account both performance and cost, while ensuring tenant constraints are not exceeded. It takes advantage of provider pricing diversity by selecting the most cost-effective provider for each query execution, thereby reducing overall costs while preserving performance. The main contributions of this paper are as follows:

- We identify and formalize the challenge of tenant-centric data replication in multi-cloud environments.
- We propose a tenant budget-aware replication strategy (TCDRM) that integrates tenant budget, popularity and response time thresholds.
- We develop a heuristic for replica placement that narrows the search space and supports cost-efficient decision-making.
- We extend CloudSim (Calheiros et al., 2011) to simulate a multi-cloud environment with integrated replication, cost modeling, and query routing.
- We validate the effectiveness of our strategy through simulations that demonstrate its ability to maintain performance while respecting tenant budgets.

The remainder of this paper is structured as follows: Section 2 reviews related work on data replication strategies in cloud and multi-cloud systems. Section 3 introduces the architecture of the proposed TCDRM strategy. Section 4 details the research methodology. Section 5 describes the simulation setup, outlines the modifications made to CloudSim, and presents the experimental results. Finally, Section 6 concludes the paper and outlines directions for future work.

## 2. Related work

One of the objectives of the data replication strategy is to satisfy specific tenant objectives, such as improving performance (Arar et al.,2024), guaranteeing data availability (Wei et al., 2010), load balancing (Edwin et al. , 2019), and reducing response time (Tos et al., 2018; Khelifa et al., 2021), reliability (Li et al., 2017), security (Ali et al., 2018), and energy efficiency (Alghamdi et al ., 2017; Seguela et al., 2019; Seguela et al, 2022), or taking into account multi-objective criteria for tenants (Miloudi et al., 2020; Boru et al., 2015, Wang & Shannigrahi, 2025; Edwin et al., 2019; Mansouri and Javidi, 2018; Mokadem and Hameurlain, 2020).

Concurrently, economic aspects, particularly profit maximization, are integrated into these strategies. In this context, Mokadem et al. (2022) propose a twofold classification based on economic orientation: provider-oriented strategies and tenant-oriented strategies. Providers-oriented replication strategies prioritize cost reduction for providers while adhering to the stipulated SLO. This approach has been illustrated in several studies, including Wei et al. (2010) and Bonvin et al. (2010), Sousa and Machado (2012), Tos et al. (2016), Liu et al. (2019), Tos et al. (2018) and Khelifa et al. (2022). These studies highlight the mechanisms for optimizing resources for the benefit of the provider, while maintaining an adequate level of service for users.

Conversely, tenant-oriented strategies prioritize the reduction of monetary costs incurred by tenants over providers; notable works in this area include those by Sakr et al. (2011) and Sharma et al. (2011), Sakr and Liu (2012), Zhao et al. (2015), Limam et al. (2019), and John and Mirnalinee (2020). These strategies aim to minimize tenant expenditure while ensuring adequate performance.

Sakr and Liu (2012) present an approach by controlling the tenant's monetary outlay on cloud database management based on SLAs. They proposed a framework that acts as a middleware between applications and cloud databases. This framework enables the dynamic, adaptive provisioning of resources according to application-defined policies to guarantee performance while minimizing monetary costs. To adapt to variations in workload, replicas are added during periods of high demand and removed when the load decreases, ensuring the continuous optimization of expenditure.

Limam et al. (2019) proposed a Dynamic Replication strategy, DRAPP (dynamic replication strategy for Availability, Performance, and Profit), which aims to meet the availability and performance requirements of cloud systems for databases while considering tenant budget and vendor profit. This approach triggers the creation of replicas when there is an SLA violation, particularly in the response time, and the cost of replication does not exceed the initial threshold budget.

Zhao et al (2015) propose a framework-based approach for optimized cost management via dynamic, intelligent resource provisioning based on application SLA requirements. This framework adjusts resources in real time and optimizes the use of database replicas to minimize costs while guaranteeing optimal performance. It relies on virtualization-based replication mechanisms, facilitating more efficient allocation of cloud resources. Database replicas are deployed on virtual machines (VMs) located in different geographical zones, enabling us to: (i) reduce costs by exploiting cost-effective cloud instances depending on the region and (ii) avoid over-provisioning costs by dynamically adjusting the number of replicas according to demand.

In this context, TCDRM is viewed as a tenant-oriented strategy. This approach triggers data replication only when the budget allocated by the tenant to a given query is exceeded. In this way, tenants' financial resources are preserved and access to data is guaranteed in line with their needs.

However, most replication strategies mentioned were designed for single-cloud environments. These approaches generally fail to consider the diversity of pricing options and configurations available in a multi-cloud environment, thereby limiting their ability to fully exploit the economic and technical advantages offered by a distributed infrastructure.

A limited number of studies have looked at how tenants' budgets are considered in a multi-cloud context. These include the contributions of Chang et al. (2012) and Abu-Libdeh et al. (2010) and Bessani et al. (2011, 2013), Chen et al. (2014) and Wu et al. (2013), Abouzamazem and Ezhilchelvan (2013), Grozev and Buyya (2015), Liu and Shen (2017), Mansouri and Buyya (2019). These studies explore mechanisms for taking advantage of pricing differences and technical specificities between multiple cloud providers while respecting tenants' budget constraints.

TCDRM is viewed as a tenant-oriented strategy. This approach triggers data replication only when the budget allocated by the tenant to a given query is exceeded. In this way, tenants' financial resources are preserved and access to data is guaranteed in line with their needs.

Chang et al (2012) propose a dynamic programming algorithm to optimize data replication across multi-cloud, in order to improve cloud storage availability. Their approach is based on the selection of providers according to two main criteria: the cost and probability of failure. Using optimization algorithms, they maximized data durability within a predefined budget. Assuming that supplier failures are independent, this method guarantees users continued access to their data, even in the event of supplier failure, while optimizing the management of replication costs. By diversifying suppliers and prioritizing those offering the best cost-availability ratio, users can minimize their expenses while ensuring the security and accessibility of their data.

Liu and Shen (2017) propose DAR (Data Storage and Request Allocation and Resource Reservation) to optimize cloud storage, a system that enables tenants to minimize their costs while respecting SLOs in a multi-cloud environment. DAR is based on two heuristic solutions. First, the dominant cost allocation algorithm identifies the dominant cost (storage, Get, or Put) for each data element and allocates it to the data center offering the lowest unit price for this dominant cost. The aim is to reduce costs within a "pay-as-you-go" framework. On the other hand, the optimal resource reservation algorithm maximizes the savings achieved through advance reservations compared to pay-as-you-go. This avoids both overestimation (overbooking) and underestimation (underbooking).

Wang et al, (2020) propose an adaptive data placement architecture, called ADP (Adaptive Data Placement Architecture) designed for mutli-cloudenvironments. This approach aims to minimize the total costs and maximize data availability by dynamically adjusting the data placement scheme according to the frequency of data access, which varies over time. ADPA combines an access frequency prediction module based on LSTM to anticipate future needs, and a placement optimization module based on reinforcement learning (Q-learning) to determine the best storage strategy as in (Najjar et al., 2025).

Aldailamy et al. (2024) propose two online algorithms: DTS (Deterministic Time Slot) and RTS (Randomized Time Slot) to dynamically optimize data replication and object placement in a multi-cloud environment, specifically targeting Online Social Networks (OSNs). These algorithms rely on real-time analysis of object popularity based on regional access and engagement rates (likes, comments, shares). Their goal is to guarantee access latency below a 250 ms threshold while minimizing inter-cloud bandwidth, storage, and replication costs. Although effective, these algorithms do not consider the variability of pricing policies across cloud providers, which may limit the economic benefits for tenants in a multi-cloud context.

These multi-cloud approaches thus offer increased flexibility and cost optimization while meeting data performance and availability requirements. However, most of these strategies are aimed at storage. To the best of our knowledge, no work has yet proposed tenant-oriented data replication strategies for databases in a multi-cloud environment.

## 3. The proposed TCDRM Strategy

The proposed TCDRM strategy aims to ensure SLA satisfaction by meeting the response-time objectives while maintaining the tenant's budget. To achieve this, our approach addressed four key questions related to data replication, each associated with a specific decision-making module within the system: (i) When and what to replicate? This module determines the optimal timing for replication and identifies the data to be replicated based on its popularity and cost. (ii) Where to replicate? handled by the placement algorithm, which selects providers based on trade-offs between cost and response time, and (iii) replica deletion managed by the deletion algorithm, which continuously monitors replica popularity to avoid unnecessary storage costs for tenants. These interconnected components form a coherent and adaptive system capable of balancing performance with cost efficiency. In addition, we introduce an economic model that estimates the query execution costs (CPU, I/O, and bandwidth), enabling the dynamic calculation of a budget threshold for replication. This cost-aware model guides replication decisions to meet financial constraints, while ensuring system responsiveness.

## 3.1. Architecture overview

We consider that data sources are distributed across multiple clouds connected to the Internet, which is a network characterized by low bandwidth and high latency. The architecture we adopt is based on middleware (Fig. 1), which serves as an interface between the user and several cloud service providers. When a query is issued, the middleware collects information from the various clouds and selects the best option for the tenant. Furthermore, each provider adopted its own pricing policy. For tenants, communication costs are often higher than storage costs. Additionally, the communication charges between different regions are higher than those within the same region, which in turn are higher than the costs of communication within the same data center.

In a multi-cloud system, each cloud provider (CP) has multiple data centers (DC) distributed across various geographic regions (RG).

Let CP = {CP1, …, CPj, …,CPr} be a set of r providers connected via the internet to the user through middleware.

Let RGp = {RGp1, …, RGpi, …, RGpq} be a set of q regions, of provider p, geographical regions connected via the internet without direct links between them. Each region includes heterogeneous data centers.

Denote DC = {DCpi1, …, DCpij, …, DCpin} to designate a set of n DCs in a region RGpi of a provider p. The bandwidth between these DCs is more abundant and cheaper compared to the first level. In these data centers, virtual machines are offered to tenants to provide storage, computing, and bandwidth services.

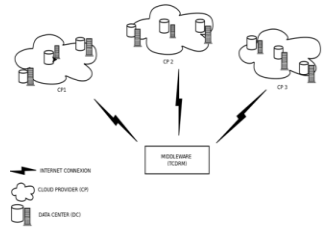

Fig.1: Considered architecture

## 3.2. Which data to replicate and when?

In the proposed strategy, the creation of a new replica depends on three main criteria: the actual query response time, economic cost associated with query execution, and popularity of the accessed data.

A new replica creation is considered only if (i) the query response time ($t_Q$) exceeds a predefined SLA threshold ($T_{SLA}$), or (ii) the monetary cost of the query ($c_Q$) (Algorithm 1, line 4), which we present later, exceeds the threshold budget ($C_{SLA}$). Additionally, the creation of replicas is conditioned by the popularity of the accessed data, which must exceed a fixed popularity threshold ($P_{SLA}$).

We used the data popularity metric proposed by Chettaoui and Ben Charrada (2012) to estimate the average request rate of a dataset over its actual usage period, starting from the time of its first access (Lazeb et al., 2021). It is defined as:

$$pdi = (\#Requests)/(T\_current - T\_(first\_request) + 1) \text{ (Equation 1)}$$

Where:
- #Requests is the total number of requests received from the dataset.
- T_current is the current timestamp (i.e., the time at which popularity is evaluated).

- T_first_request is the timestamp for the first recorded access to the dataset.

This approach avoids the bias introduced by inactive periods before the first use of the dataset and provides a more realistic representation of its popularity during its active lifetime.

When a query Q is submitted, it is executed immediately according to the execution plan provided by the DBMS. Let us assume that this Query Execution Plan is provided for each tenant query at the end of execution. The query's response time and monetary cost, calculated based on the resources used, were then compared with the thresholds specified in the SLA. We relied on the strategy of Tos et al. (2016) to estimate the response time (Algorithm 1, line 2). If any of these thresholds are exceeded, a detailed analysis of the query is performed by examining all data necessary for its execution. If the popularity of the data ($P_{SLA}$) exceeds a predefined threshold in the SLA (Algorithm 1, Line 7), the creation of replicas is considered.

**Algorithm 1: Replica Creation.**

INPUTS:

    Q:   Query to be executed,
    $T_{SLA}$: Response Time Threshold,
    $C_{SLA}$: Monetary Cost Threshold,
    D:   Set of data items,
    $P_{SLA}$: Data Popularity Threshold

VARIABLE:

    $t_Q$: Response Time of query Q,
    $c_Q$: Monetary Cost of query Q,
    pdi: Popularity of data item di

OUTPUT:

  RD: List of data items to replicate (initialilly empty)

BEGIN

1. Execute Q
2. $t_Q$ ← responseTime(Q) // Calculate response time of Q
3. $c_Q$ ← monetaryCost(Q) // Calculate monetary cost of Q
4. IF (tq > $T_{SLA}$ OR cq > $C_{SLA}$) THEN
5.    FOR each data di in D DO
6.      pdi ← dataPopularity(di) // Compute popularity of data item
7.      IF (pdi > $P_{SLA}$) THEN
8.        Add di to RD //Mark for replication
9.      END IF
10.   END FOR
11. END IF

END

### 3.3. Where to replicate?

TCDRM is a heuristic-based algorithm for replica placement designed to reduce the search space and minimize the time required to identify suitable providers for hosting new replicas. The search is constrained to providers offering hosting rates below the tenant's predefined budget threshold ($C_{SLA}$) while also ensuring that the expected response time remains within the SLA-defined limit ($T_{SLA}$).

    Upon the execution of query Q and identification of the relevant data (RD) for replication, the placement algorithm (Algorithm 2, lines 1–2) evaluates candidate providers using simulation. For each

provider, we simulated query execution over the RD dataset to estimate two key metrics: (i) the monetary cost of query execution encompassing CPU, storage, and bandwidth usage, and (ii) the estimated response time based on the current load and resource availability. These simulations are crucial for assessing the provider suitability.

To reduce the search space, the algorithm first filters out providers whose estimated costs exceed the $C_{SLA}$ (Algorithm 2, line 5). From the remaining candidates, it selects the first provider that meets the TSLA response time threshold (Algorithm 2, line 7). This two-step filtering process effectively narrows the search space to providers that meet both economic and performance requirements, thereby improving the overall efficiency of the replica placement strategy.

**Algorithm 1: Replica placement heuristic.**

INPUTS:
  RD: A list of data items identified for replication,
  P: A list of available cloud providers,
  Q: A query that triggered the replication decision,
  $T_{SLA}$: The response time threshold defined in the SLA,
  $C_{SLA}$: The maximum monetary cost allowed for query execution, based on the tenant's budget

OUTPUT:
  The final replica placement decision, i.e., the selected provider(s) for each data item.

```
BEGIN
1.  FOR each data_to_replicate rk in RD DO
2.      pi ← firstProvider(P)
3.      WHILE pi AND (eMci > CSLA OR eRespTi > TSLA) DO
4.          eMci ← estimatedMonetaryCost(Q, rk, pi)
5.          IF (eMci < CSLA) THEN
6.              eRespTi ← estimatedResponseTime(Q, rk, pi)
7.              IF (eRespTi < TSLA) THEN
8.                  PLACE(rk, pi)
9.              END IF
10.         END IF
11.         pi ← nextProvider(P)
12.     END WHILE
13. END FOR
END
```

### 3.4. Replica Deletion

To avoid unnecessary storage costs for tenants and maintain an economically sustainable replication strategy, the TCDRM framework integrates a replica deletion module. We regularly monitor the popularity of the data on servers hosting replicas. If a replica becomes less popular, meaning its usage falls below the PSLA threshold for a certain time interval deltaT, it is deleted. deltaT represents a configurable observation window expressed in evaluation periods (e.g., hours or scheduling intervals). A replica is deleted only if its popularity remains consistently below the threshold for the entire duration of the deltaT. This approach avoids frequent creation and deletion of replicas, thereby limiting unnecessary costs.

The minimum number of replicas required to ensure availability is not addressed in this proposal, because our strategy is tenant-oriented.

---

**Algorithm 3: Deletion of low-popularity replicas.**

---

INPUTS:
  PR:   A predefined observation period during which access patterns are evaluated,
  DR:   The set of currently replicated data items,
  $P_{SLA}$: The popularity threshold defined in the SLA

OUTPUT:
  Deletion of low-popularity replicas

BEGIN
  1. FOR each Period p in PR
  2.     FOR EACH ReplicatedData rd in DR
  3.         popularity ← calculatePopularity(rd)
  4.         IF popularity < $P_{SLA}$ during deltaT  THEN
  5.             deleteReplica(rd)   //Replica is deleted only if it is less popular than the threshold
  6.         ENDIF
  7.     END FOR
  8. END FOR
END

---

### 3.5. Economic Model

In a multi-cloud environment, the economic model plays a central role in the design of data management strategies. Unlike traditional pre-cloud approaches, where resources were statically provisioned, cloud providers charge for access to services (storage, bandwidth, compute) based on a pay-as-you-go model (Foster et al., 2008). Some replication strategies in multi-cloud environments incorporate economic aspects (Liu and Shen,2017; Aldailamy et al., 2024) by aiming to reduce monetary costs. However, they do not explicitly consider the tenant's budget constraint as a central parameter in their decision-making mechanisms. The proposed TCDRM in this paper fully embraces this economic logic by placing tenant budget constraints at the core of its decision-making mechanisms. The proposed replication strategy thus aims to ensure the required performance levels while controlling the costs borne by the tenant, leveraging the pricing diversity across multi-cloud providers. The economic model therefore becomes a structuring parameter both a constraint and an opportunity in optimizing placement decisions.

Thus, for each query submitted by the tenant, the total cost ($C_Q$) is calculated using the following formula:

$$C_Q = C_{CPU} + C_{IO} + C_{bandwidth} \quad \text{(Equation 2)}$$

Each component reflects a critical aspect of cloud resource usage.

- CCPU: Cost associated with the CPU time used during query execution. The CPU cost is commonly measured in vCPU-seconds, with typical pricing ranging from $0.000011 to $0.000060 per vCPU-second, depending on the provider and region.
- CIO: Represents the cost of storage I/O operations, including reading from and writing to disks and the cost of creating new replicas. Providers such as AWS and Azure charge per 1,000 operations or per GB transferred to or from storage.

- Cbandwidth: Refers to the cost of transferring data across regions or clouds. This is a significant factor in mutli-cloud environments. For example, the outbound data transfer may range from $0.01 to $0.12 per GB.

These cost parameters were not arbitrarily chosen. These are based on publicly available pricing from major cloud providers (e.g., AWS, Azure, and Google Cloud), which are periodically updated. Our model uses configurable parameters that can be adapted to reflect the current pricing and tenant preferences.

To enforce tenant cost control, we introduce a Monetary Cost Threshold ($C_{SLA}$) that defines the maximum acceptable cost for executing a query. When a tenant submits query Q, TCDRM selects the provider whose estimated cost is within the $C_{SLA}$ limit. If multiple providers qualify, the provider offering the lowest level is chosen. This decision mechanism abstracts the provider operations as a "black box" and focuses entirely on optimizing tenant-side cost efficiency.

## 4. Experimental analysis

In this section, we evaluate the proposed TCDRM strategy. First, we present the simulation tool and the various experiments conducted, including the generation of queries. Then, we compare the performance of TCDRM against NoRepLc (No-Replication-Less cost), a baseline strategy without replication that integrates a basic cost-aware mechanism by selecting the cheapest provider for each query in a multi-cloud environment. Finally, we conclude with an analysis of the results.

The fundamental assumptions of our simulation were as follows:
- The TCDRM strategy is tenant-centric. Cloud service providers are treated as black boxes. In other words, we do not consider intra-DC or inter-DC communication within a single provider. Instead, we focus on the interactions between different regions (interRegion) to ensure the validity of the experiment.
- For each query, we consider the submission region, relationships necessary for processing, and selection of virtual machines (VMs) responsible for execution.
- During the execution of each query, we always selected the least expensive virtual machine, including in the experiments conducted with NoRepLc.

### 4.1. Simulation Environment

To evaluate the proposed TCDRM strategy, we used CloudSim (Calheiros et al., 2011), a widely adopted simulator in cloud computing research. However, since CloudSim was originally designed for single-cloud scenarios, it lacked support for data replication, distributed database query processing, and cost-aware multi-cloud environments. Therefore, we implemented several significant extensions to meet these specific requirements.

First, CloudSim was extended to simulate a realistic multi-cloud ecosystem comprising multiple providers such as AWS, Google Cloud, and Microsoft Azure. Each provider is modeled with its own data centers, virtual machine types, pricing schemes, and performance parameters. The data centers are geographically distributed across Europe, the United States, and Asia. Each provider can be configured independently, enabling the simulation of heterogeneous environments in terms of resource availability, cost structures, and service quality.

Second, the default cloudlet model was enhanced to support interdependent task execution, which is essential for accurately simulating complex database queries composed of multiple related operations distributed across providers.

Third, a hierarchical network latency model was introduced, covering various levels of communication: intra-VM, intra-datacenter, inter-datacenter (within the same provider), and inter-cloud (between different providers). Bandwidth between nodes is also configurable, affecting both execution time and data transfer costs.

Fourth, an economic model was integrated to capture resource usage costs, including CPU, I/O, and bandwidth (as detailed in Section 4.4), as well as intra- and inter-provider data transfer expenses. This module relies on realistic pricing models inspired by major cloud providers (cf Tab. 1), enabling cost-sensitive simulations and supporting budget-aware replication strategies.

Finally, a series of simulations was conducted using 1,000 queries, each involving three to six relationships distributed across multiple providers. Additionally, 20 relationships were placed in each region per provider to represent a realistic multi-cloud data distribution scenario.

### 4.2. Definition of queries

We considered queries of varying complexity, based on the number of joins and placement of relations: simple queries and complex queries.

Simple queries perform joins by associating one relationship with each of the three regions.

However, complex queries execute joins between relationships located across all three regions, with at least two relationships per region.

The region of origin for the queries was selected randomly at the time of submission.

Tab. 1: Configuration parameters.

|  | Google | | | AWS | | | AZURE | | | Parameter | Value | |
|---|---|---|---|---|---|---|---|---|---|---|---|---|
|  | US | UE | AS | US | UE | AS | US | UE | AS | Number of Providers | 3 | |
| CPU ($ / 10⁶ MI) | 0.020 | 0.025 | 0.027 | 0.020 | 0.018 | 0.020 | 0.0095 | 0.0090 | 0.0080 | Number of Regions within a Provider | 3 | |
| I/O ($ / GB) | 0.006 | 0.006 | 0.006 | 0.0096 | 0.008 | 0.0096 | 0.0120 | 0.0096 | 0.0090 | Number of VMs within a Region | 20 | |
|  |  |  |  |  |  |  |  |  |  | Average (Avg) size of a relation | 450 MB | |
|  |  |  |  |  |  |  |  |  |  | $T_{SLA}$ | 200 ms | Simple Queries |
| BW IntraDC ($ / GB) | 0.0015 | 0.002 | 0.004 | 0.0015 | 0.002 | 0.004 | 0.0015 | 0.002 | 0.004 | $C_{SLA}$ | 0.015 $ per query | |
| BW InterRegion ($ / GB) | 0.008 | | | 0.008 | | | 0.008 | | | $T_{SLA}$ | 400 ms | Complex Queries |
|  |  |  |  |  |  |  |  |  |  | $C_{SLA}$ | 0.040 $ per query | |
| BW Inter-Provider ($ / GB) | 0.01 | | | 0.01 | | | 0.01 | | | $P_{SEUIL}$ | 200 | |

We distributed the relationships across various providers and executed the same query 1,000 times analyze the impact of our method.

### 4.3. Simulation Results

To validate the proposed strategy, several metrics were measured during the simulation. We focused on the following aspects:
(i) The replication factor,
(ii) The impact on response time, and
(iii) Impact on costs.

#### 4.3.1. Replica Factor

The goal of TCDRM is to create new replicas, when necessary, to reduce client costs. Although the creation of these replicas incurs additional costs, they should be viewed as long-term investment. Fig. 1 shows the variation in the number of replicas as a function of the number of queries. During different experiments, we set the data popularity threshold (PSLA) to 200, representing the frequency of access to the relationship. As such, the number of replicas gradually increases from the threshold onward, as replica creation is conditioned by data popularity and other predefined thresholds in the SLA, such as the response time and the maximum allowable query cost.

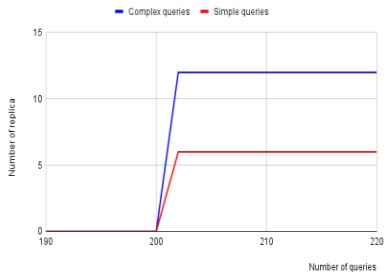

Fig. 2 : Replica Factor: Number of replicas created as a function of the number of executed queries.

Fig. 2 shows the number of replicas generated as a function of the number of queries. We observe an increase in the number of replicas as a function of the query complexity. The maximum number of replicas differs for each type of query because the number of joins and the placement of each relationship vary depending on the query type. TCDRM generates replicas in a manner tailored to needs by analyzing each query.

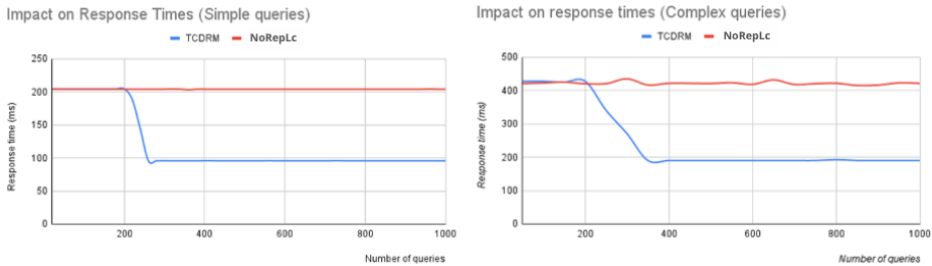

Fig. 3 : Impact on Response Time: Average response time as a function of the number of queries, comparing simple and complex query workloads.

Fig. 3 shows the average response time as a function of the number of queries for both simple and complex queries. For both query types, we observed that the trends of the curves representing TCDRM and NoRepLc were similar up to a certain number of queries. However, the curve representing TCDRM did not follow a linear progression. A decrease was observed up to a certain point, followed by stabilization. Starting from the PSLA, we observe a decline in the curve for both query types as TCDRM begins to replicate, and the number of inter-provider exchanges gradually decreases until the maximum number of replicas is reached.

With NoRepLc, each request requires data transfer between providers (inter-provider data transfer), which increases response times. Because the same query is repeated multiple times, an almost linear curve is created, as shown in Fig. 2. However, we still observed fluctuations in the NoRepLc curve for complex queries. This is because the origin of the query was random for each query. In addition to inter-provider exchanges, these fluctuations represent exchanges between regions, as relationships may be located in different regions.

In other words, the curves show a significant reduction in the response times for TCDRM. Although optimizing the response times is not the primary goal of our strategy, TCDRM significantly reduces the response time compared to NoRepLc. By utilizing the already created replicas, the bandwidth consumption is reduced, which in turn halves the response time. We expect these gains to become more significant as the number of providers increases.

### 4.3.2. Effect on Bandwidth Consumption

In this experiment, we evaluated the cost associated with the bandwidth consumption. Although each provider sets its own pricing structure, a common factor emerges: the high cost associated with data transfer between different providers (interProvider). Therefore, these transfers can have a significant impact on the final costs borne by tenants.

Fig. 4 presents the bandwidth consumption between providers (interProvider) and between regions (interRegion) for both NoRepLc and TCDRM for both simple and complex queries. It is important to note that inter-region exchanges occur within the same provider.

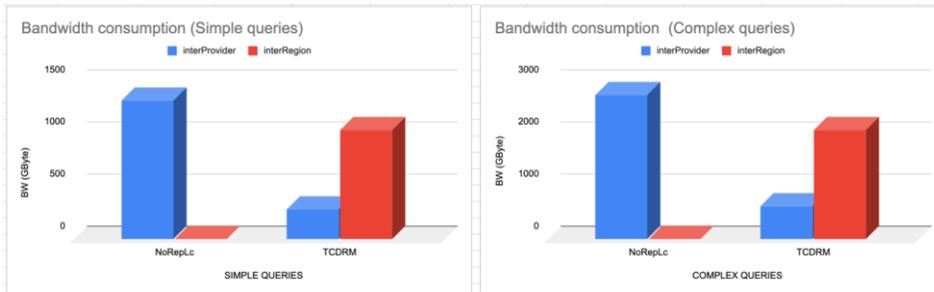

Fig. 4 : Bandwidth Consumption Comparison: Inter-rovider vs. interRegion bandwidth usage under TCDRM and NoRepLc.

In Fig. 4, we observe a similarity in the behavior of the graphs, although their scales differ. In the case of NoRepLc, the bandwidth consumption between providers (inter-providers) is higher than internal consumption. This disparity is explained by the distribution of relationships across various providers, forcing the virtual machine to execute a query to retrieve these relationships. By contrast, TCDRM creates replicas of the relationships, favoring transfers primarily within the same provider once the replicas are created. As a result, replication eliminates inter-provider exchanges, and TCDRM effectively satisfies this requirement.

Fig. 5 illustrates the cost of bandwidth consumption per query for both NoRepLc and the TCDRM. These results indicate that the bandwidth consumption cost of NoRepLc remains stable and high. However, the bandwidth consumption costs for TCDRM decrease over time. The number of replicas and their placements have a significant effect on bandwidth consumption.

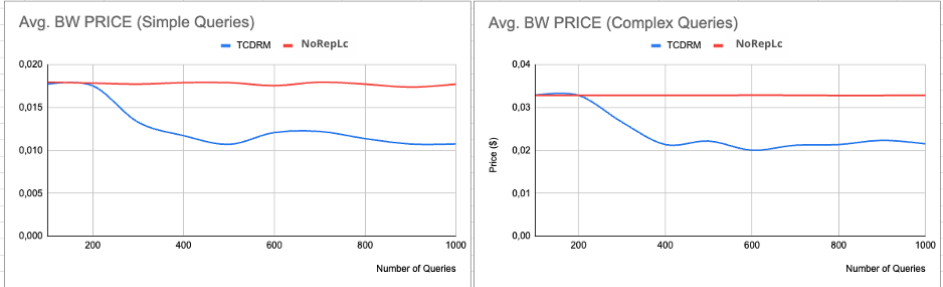

Fig. 5 : Bandwidth Cost per Query: Comparison of bandwidth consumption cost per query between NoRepLc and TCDRM for Simple and Complex Queries

In the TCDRM curve, we observed an oscillation before this decline. This observation can be explained by the fact that replica creation occurs during these intervals, and their usage gradually increases, even when processing the same query. Additionally, as the curve reaches its lowest point, it does not follow a linear trajectory owing to the persistence of inter region exchanges. Indeed, the heuristic (cf., Section 4.2) seeks an optimal placement to reduce the search space, and the replicas of the relationships created may end up in regions different from the origin of the query.

Fig. 6 shows the cumulative price of bandwidth usage obtained by experimenting with NoRepLc and TCDRM for simple and complex queries.

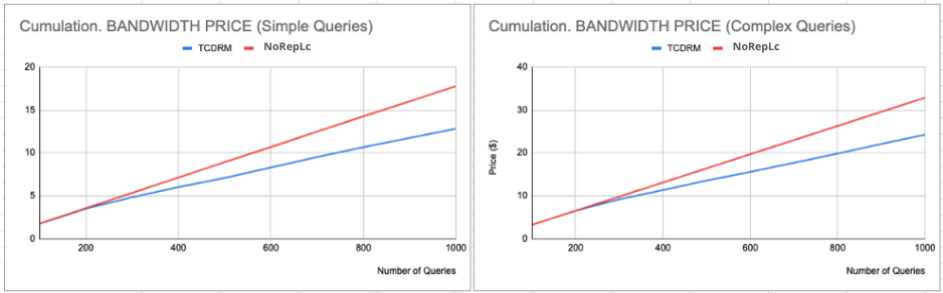

Fig. 6 : Cumulative Bandwidth Cost: Cumulative cost as a function of the number of executed queries.

It was observed that with NoRepLc, a stable trajectory was achieved, whereas with TCDRM, the trajectory gradually diverged over time. The constant curve of NoRepLc indicates continuous price stability without replication, whereas the increase is less significant for TCDRM. This highlights the clear advantage of TCDRM in terms of bandwidth costs for the tenant depending on the number of queries made.

### 4.3.3. Effect on Total Price

Fig. 7 illustrates the comparison of the total costs between NoRepLc and TCDRM for both simple and complex queries. This highlights the expenses associated with storage, bandwidth, and CPU. Overall, the bandwidth cost was significantly higher than those of the others. The CPU cost remained constant for all query types and strategies. The storage cost is nonexistent for NoRepLc and negligible for TCDRM.

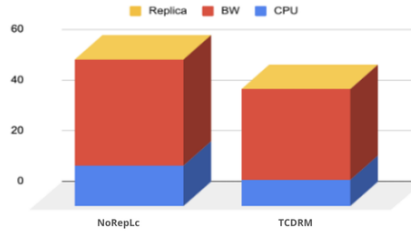

Fig. 7 : Total Price: comparison of the total costs between NoRepLc and TCDRM

Our strategy aims to adhere to the client's initial budget by ensuring that the price paid by the tenant does not exceed a set threshold, while maintaining a reasonable response time. Fig. 7 shows that this objective was achieved by comparing the total cost obtained with the NoRepLc. Bandwidth cost plays a predominant role in the total cost. The high costs associated with data transfer have a significant impact on the total cost, while the cost of storing replicas remains negligible.

### 4.4. Result Analysis

The experimental results clearly support the original goal of TCDRM: to deliver a tenant-centric, budget-aware data replication framework tailored for complex multi-cloud environments. A key strength of TCDRM lies in its ability to strategically balance storage costs, data transfer overhead, and budgetary constraints, addressing cost dimensions that are often overlooked in conventional replication mechanisms. In particular, the cost of creating new replicas includes both inter-provider data transfer and storage expenses. Based on average pricing from major cloud providers such as AWS, Azure, and GCP, standard storage is approximately $0.02/GB per month, while inter-cloud data transfer (egress) averages $0.10/GB. Thus, transfer costs can be five times higher than storage costs, emphasizing the importance of reducing unnecessary inter-cloud transfers an objective that TCDRM systematically pursues.

The system was evaluated on a workload of 1,000 queries, a size chosen to strike a balance between realism and analytical clarity. While it does not simulate all real-world scales, this setup is sufficient to highlight key performance trends, illustrate cost amortization effects, and demonstrate consistent budget compliance across diverse access patterns. These experiments serve as a proof of concept, validating the soundness and benefits of TCDRM's core principles. Future work will extend the evaluation to larger workloads to confirm long-term scalability and robustness.

TCDRM further shows strong adaptability to different workload profiles. It performs especially well under repetitive or clustered access patterns, where the cost of replication is quickly amortized through repeated data reuse. Even in more irregular or unpredictable workloads, the framework continues to meet performance goals while remaining budget-compliant, underscoring its flexibility and practical value.

### 5. Conclusion

This paper introduces TCDRM, a novel tenant-centric data replication strategy specifically designed for multi-cloud environments. Unlike previous strategies that primarily optimize for provider profits or are limited to single-cloud scenarios, TCDRM prioritizes tenant budget constraints while maintaining acceptable performance levels. By dynamically analyzing query execution metrics against predefined SLA thresholds for response time, cost, and data popularity, our strategy creates strategically placed replicas that leverage the diverse pricing structures of multiple cloud providers.

Our comprehensive evaluation through CloudSim simulations demonstrates that TCDRM significantly outperforms non-replicated approaches in both economic and performance dimensions. Specifically,

TCDRM reduces bandwidth consumption costs by up to 78% for complex queries while simultaneously decreasing response times by an average of 51%. These improvements are achieved through intelligent replica placement that minimizes expensive inter- provider data transfers and leverages lower-cost intra-provider communication.

The findings from this research have important implications for organizations deploying data-intensive applications across multiple clouds. TCDRM provides a practical framework for balancing the cost-performance trade-off in these complex environments, particularly benefiting tenants with strict budget constraints. The threshold-based approach also offers flexibility in adapting to different workload characteristics and tenant requirements.

A known limitation of the current approach is the reliance on static thresholds to guide replication decisions. In dynamic environments where query patterns, data popularity, and pricing may fluctuate—fixed thresholds may lead to suboptimal decisions. To overcome this, we plan to incorporate machine learning techniques (Najjar et al., 2024) capable of dynamically adjusting these thresholds in real time, based on observed workload characteristics, data popularity trends, and evolving cost structures.

Future work will focus on three key directions: (1) comparing TCDRM with other state-of-the-art replication strategies to further validate its advantages, (2) implementing and testing the approach in real multi-cloud environments to confirm the simulation results, and (3) extending the framework to address data consistency and security challenges that arise in multi-cloud replication scenarios. Additionally, we plan to investigate machine learning approaches for dynamically adjusting thresholds based on changing workload patterns, provider pricing structures and data popularity predictions.